\documentstyle[11pt,newpasp,twoside]{article}
\def\edcomment#1{\iffalse\marginpar{\raggedright\sl#1\/}\else\relax\fi}
\reversemarginpar

\begin{document}
\title{Probing Baryons in Galactic Halos and Gas Near Galaxies}
\author{Kenneth R. Sembach}
\affil{Space Telescope Science Institute, 3700 San Martin Dr., Baltimore,
MD 21218}
\author{Bart P. Wakker, Blair D. Savage, Philipp Richter, \& Marilyn Meade}
\affil{Astronomy Dept. University of Wisconsin-Madison, 475 N.Charter St.,
Madison, WI 53706}

\begin{abstract}
 We describe an extensive
FUSE survey of highly 
ionized oxygen in the vicinity of the Milky Way that serves as an example of 
the type of study that would be desirable for other galactic systems.
Understanding the origin of 
hot gas in the vicinity of galaxies and its relationship to the intergalactic
medium presents a major observational challenge.  \hbox{Ultraviolet}
absorption-line spectroscopy is currently 
the most direct means for comprehensive investigations of the gas 
in galactic environments, but even with present (and near-term) facilities 
the number of background objects available to probe nearby galaxy halos and 
low-redshift cosmological structures is limited. Studying 
these structures over a range of impact parameters and angular separations 
would provide fundamental information about the baryonic content of the hot
gas, its physical conditions, and its origins.  A large space
telescope optimized for high resolution spectroscopy in the 
900--3200\,\AA\
wavelength region at a sensitivity sufficient to observe faint AGNs/QSOs 
at angular separations of $< 1\deg$  would be ideal for such studies.
\end{abstract}

\section{Introduction}
Understanding galaxy formation and evolution requires observational 
information about the hot, highly-ionized gas found in and near galaxies.  
A complete picture of the relationship between 
hot gas in galaxies and the intergalactic medium (IGM) does not
yet exist because a variety of processes affect the 
heating and distribution of the interstellar gas in and
around galaxies.  Galaxy formation, accretion of satellite 
galaxies, tidal interactions, star-formation, galactic winds, and galaxy-IGM 
interactions may all contribute to the production of the hot gas observed
in the IGM and extended galactic halos.

In this article, we outline a program we have begun with the Far 
Ultraviolet Spectroscopic Explorer (FUSE) to study the hot
gas in the vicinity of the Milky Way.  The study is described in detail 
in a series of three articles devoted to probing the highly ionized 
oxygen (O\,VI) absorption along complete paths through the Galactic 
halo and Local Group.  The articles include a catalog of the spectra and basic 
observational information (Wakker et al. 2002), a study of the hot
gas in the Milky Way halo (Savage et al. 2002b), and an investigation of the
highly ionized high velocity gas in the vicinity of the Galaxy (Sembach
et al. 2002).  Here, we summarize the high velocity gas results and comment
on their implications for understanding other types of highly ionized gas.
This comprehensive study serves as a prime example of 
what can be learned if serious efforts are made to conduct large 
spectroscopic observing programs with existing space-based instrumentation.
Similar types of studies conducted with future generations 
of instruments on the Hubble Space Telescope (HST) and other large space 
telescopes will enable us to explore the properties of gas in other groups of 
galaxies in much greater detail than is presently possible.

In Table~1 we list basic information for some of the best ultraviolet (UV)
absorption-line diagnostics of hot gas.  The table 
entries include the atom or ion, rest wavelength of the transition, 
cosmological redshift required 
to shift the line(s) to an observed wavelength of 1150\,\AA, the temperature
at which the atom or ion peaks in abundance under conditions of 
collisional ionization equilibrium, the thermal width of the line at that
temperature, and a logarithmic relative line strength that depends 
on the cosmic abundance of the element ($A$), the line strength ($f\lambda$),
and the peak ionization fraction ($\phi$) of the ion
in collisionally ionized gas.
Larger values of $Af\lambda\phi$ indicate easier detectability.

\begin{table}
\caption{Ultraviolet Diagnostics of Hot Gas at Low Redshift}
\begin{tabular}{lccccc}
\tableline\tableline
Species & $\lambda$(\AA) & $z$(1150\,\AA) 
& $T_{CIE}$ (K)\tablenotemark{a} 
& b$_{th}$(km~s$^{-1}$)\tablenotemark{b} 
& log [$Af\lambda\phi$]\tablenotemark{c} \\
\tableline
H\,I Ly$\alpha$ & 1215.7 & --- & $<10^5$ & 40.8\tablenotemark{d} & $-2.06$\tablenotemark{d}\\
H\,I Ly$\beta$  & 1025.7 & 0.12 & $<10^5$ & 40.8\tablenotemark{d} & $-2.86$\tablenotemark{d} \\
H\,I Ly$\gamma$ & 972.5 & 0.18 & $<10^5$ & 40.8\tablenotemark{d} & $-3.32$\tablenotemark{d} \\
C\,IV & 1548.2, 1550.8 & --- & $1.0\times10^5$ & 11.8 & $-1.51, -1.81$\\
N\,V  & 1238.8, 1242.8 & --- & $1.8\times10^5$ & 14.6 & $-2.35,-2.65$\\
O\,IV & 787.7 & 0.46 & $1.6\times10^5$ & 12.9 & $-1.36$\ \\
O\,V  & 629.7 & 0.83 & $2.5\times10^5$ & 16.1 & $-0.97$\ \\
O\,VI & 1031.9, 1037.6 & 0.11 & $2.8\times10^5$ & 17.1 & $-1.65, -1.95$\ \\
Ne\,VIII & 770.4, 780.3 & 0.48 & $5.6\times10^5$ & 21.5 & $-2.36,-2.66$\\
Mg\,X & 609.8, 624.9 & 0.87 & $1.1\times10^6$ & 27.4 & $-3.31,-3.61$ \\
\tableline
\end{tabular}
\tablenotetext{a}{Temperature of maximum ionization fraction in 
collisional ionization equilibrium (Sutherland \& Dopita 1993).}
\tablenotetext{b}{Thermal line width, $b = (2kT/m)^{1/2}$, at 
$T=T_{CIE}$ unless indicated otherwise.}
\tablenotetext{c}{Values of $f\lambda$ are from Morton (1991) 
and Verner et al. (1994).  Values of $A$ (abundance relative to hydrogen
on a logarithmic scale where 
H\,=\,12.00, C\,=\,8.55, N\,=7.97, O\,=\,8.87, Ne\,=\,8.07, and 
Mg\,=\,7.59) are from Grevesse \& Noels (1993) and Anders \& Grevesse (1989). 
Values of $\phi$ are from Sutherland \& Dopita (1993).}
\tablenotetext{d}{Value at $T = 10^5$ K.}
\end{table}

The O\,VI $\lambda\lambda1031.926, 1037.617$ doublet lines
 are the best UV resonance lines to use for 
kinematical investigations of hot ($T \sim 10^5-10^6$\,K) gas in the 
low-redshift universe.  Oxygen is the most abundant element heavier than 
helium, and the O\,VI lines have large oscillator strengths.  
Lower ionization UV lines observable at high spectral resolution 
 are generally
either much weaker than the O\,VI lines (e.g., N\,V 
$\lambda1238.821, 1242.804$) or are better tracers of collisionally ionized 
gas at temperatures $T < 10^5$\,K (e.g., 
C\,IV $\lambda1548.195, 1550.770$, C\,III $\lambda977.020$,
Si\,IV $\lambda1393.755, 1402.770$, Si\,III $\lambda1206.500$).
This latter set of ions is also considerably more susceptible to 
photoionization than O\,VI.

X-ray spectroscopy of the interstellar or intergalactic gas in 
higher ionization lines (e.g., O\,VII, O\,VIII) is possible with
XMM-Newton and the Chandra X-ray Observatory for a small number 
of sight lines toward AGNs and QSOs, but the spectral resolution 
(R~$\equiv \lambda/\Delta\lambda < 1000$)
is modest compared to that afforded by FUSE (R~$\sim 15,000$).  While 
the X-ray lines provide extremely useful information about the amount of 
gas at temperatures greater than $10^6$\,K, the interpretation of where that
gas is located, or how it is related to the $10^5-10^6$\,K gas traced by
O\,VI, is hampered at low redshift by the complexity of the hot ISM and IGM
along the sight lines observed.

\section{An Example: The Milky Way and its High Velocity Cloud System}

We have conducted a study of the highly ionized high velocity gas in the 
vicinity of the Milky Way using an extensive set of FUSE data.  
We summarize the results for  the sight lines toward 100 AGNs/QSOs and two 
distant halo stars in this section (see Sembach et al. 2002; 
Wakker et al. 2002).  For the purposes of this study, gas with $|v_{LSR}|
\stackrel{_>}{_\sim} 100$ km~s$^{-1}$ is typically identified as ``high 
velocity'', while lower velocity gas is 
attributed to the Milky Way disk and halo.  A sample 
spectrum from the survey is shown in Figure~1.

\begin{figure}[h!]
\includegraphics{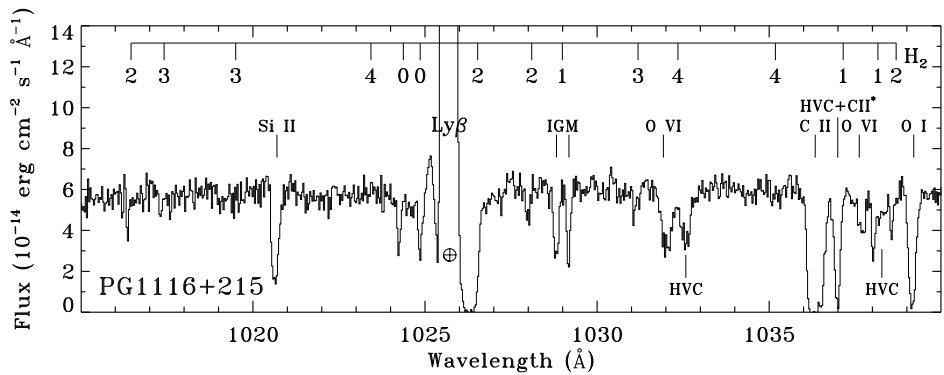}
\vspace{1.9in}
\caption{\footnotesize
A portion of the FUSE spectrum of PG\,1116+215 in the 1015--1040\,\AA\
spectral region. Interstellar and intergalactic lines are indicated.  The
numbers under the top tick marks denote the rotational levels of the H$_2$
lines ($J=0-2$ lines are prominent).  The locations of high velocity O\,VI 
features are indicated below the spectrum.}
\end{figure}

\subsection{Observational Results}

We have identified approximately 85 individual 
high velocity O\,VI features along the 102 sight lines in our sample.
A critical part of this identification process involved detailed consideration
of the absorption produced by O\,VI and other species 
(primarily H$_2$) in the thick disk and halo of the Galaxy, as well as the 
absorption produced by low-redshift intergalactic absorption lines of 
H\,I and ionized metal species. We searched for absorption in a velocity 
range of $\pm1200$ km~s$^{-1}$ centered on the O\,VI 
$\lambda1031.926$ line.  With few exceptions, the high velocity O\,VI
absorption is confined to $|v_{LSR}| \le 400$ km~s$^{-1}$,  
indicating that the O\,VI features observed are either associated with 
the Milky Way or nearby clouds within the Local Group.

\begin{figure}[h!]
\includegraphics{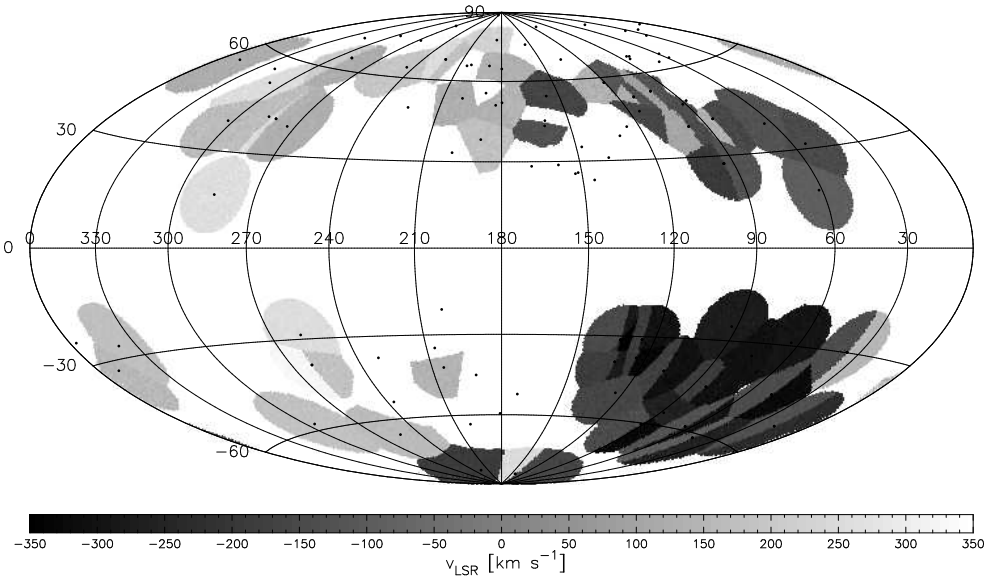}
\vspace{3.0in}
\caption{\footnotesize All-sky projection of the high velocity O\,VI 
features
in the LSR reference frame.   The Galactic anti-center is
at the center of the plot, and Galactic longitude increases to the left.
This map does not include gas at velocities 
attributed to the thick disk/halo of the Galaxy 
($|v_{LSR}| \stackrel{_<}{_\sim} 100$ km~s$^{-1}$). 
The velocities of the O\,VI features are gray-scale coded
and displayed as filled regions of radius 12$^\circ$. 
When two features
are detected within 12$^\circ$ 
 of each other (either along the same sight line 
or along adjoining sight lines), the shaded area size is adjusted accordingly.
Points with no shading indicate directions where no high velocity O\,VI is
observed.}
\end{figure}

The high velocity O\,VI features have velocity centroids ranging 
from $-372 < v_{LSR} < -90$ km~s$^{-1}$ to 
$+93 < v_{LSR}
< +385$ km~s$^{-1}$. There are an additional 6  
confirmed or very likely ($>90$\% confidence) detections and 2 tentative 
detections of O\,VI 
between $v_{LSR} = +500$ and +1200 km~s$^{-1}$; these very high velocity
features probably trace intergalactic gas beyond the Local Group.  We plot
the sky distribution of the high velocity O\,VI 
in Figure~2, where we code the velocities
of the features by shading circular regions around the sight lines for which
high velocity O\,VI is detected.

Most of the high velocity O\,VI features have velocities incompatible 
with those of Galactic rotation (by definition).  The dispersion about the 
mean of the high velocity O\,VI
centroids  decreases when the velocities are converted from the
Local Standard of Rest (LSR) into the Galactic Standard of Rest (GSR) and 
the Local Group Standard of Rest (LGSR) reference frames.   While this 
reduction is expected if the 
O\,VI is associated with gas in a highly extended Galactic corona or 
in the Local Group,  it does not provide sufficient proof by itself of an
extragalactic location for the high velocity gas.  Additional information,
such as the gas metallicity or ionization state, is needed to constrain the 
cloud locations.

The line widths of the high velocity O\,VI features range from
$\sim$16 km~s$^{-1}$ to $\sim$81 km~s$^{-1}$, with an average of 
$\langle {\rm b} \rangle = 40\pm14$ km~s$^{-1}$.
The lowest values of b are close to the thermal width of 17.1 km~s$^{-1}$
expected for O\,VI
at $T = 2.8\times10^5$\,K, while higher values of b 
require additional non-thermal broadening mechanisms or gas
temperatures significantly larger than $2.8\times10^5$\,K.

The high velocity O\,VI features have logarithmic column densities
of 13.06 to 14.59, with an average of $\langle \log N \rangle = 
13.95\pm0.34$ and a median of 13.97.  We show the distribution of the 
high velocity O\,VI column densities on the sky in Figure~3.  
The average high velocity O\,VI column density is a factor of 2.7 times lower 
than the typical low velocity O\,VI column density found for the same
sight lines 
through the thick disk/halo of the Galaxy (see Savage et al. 2002b).

\begin{figure}[h!]
\includegraphics{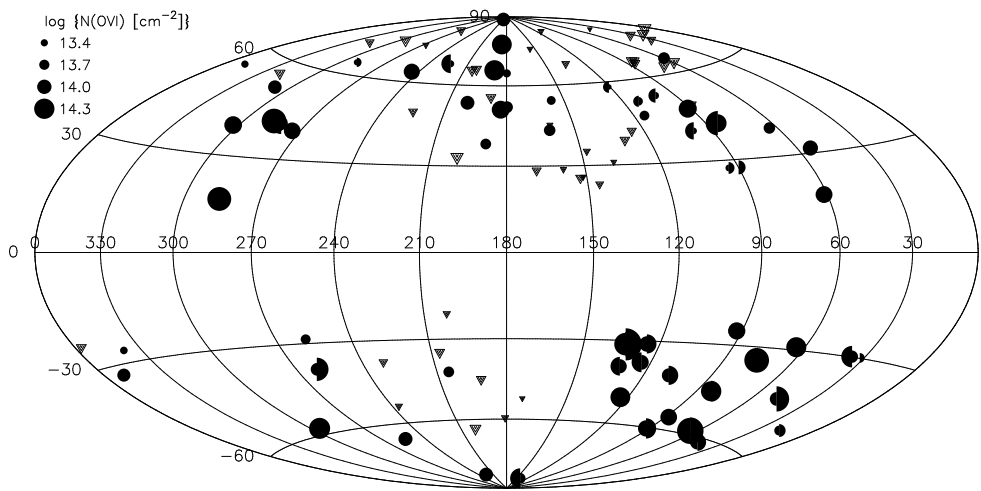}
\vspace{2.7in}
\caption{\footnotesize
All-sky projection of the high velocity O\,VI
column densities observed in our FUSE survey. 
The logarithmic column density is coded according to symbol 
size (see legend), with the symbol split 
in half if two features are present.  Triangles indicate upper limits.
Note that large column densities of high velocity O\,VI ($\log N > 14$) 
are found in many regions of 
the sky.}
\end{figure}

We detect high velocity O\,VI $\lambda1031.926$ absorption 
with total values of $W_\lambda > 30$ m\AA\ at 
$\ge 3\sigma$ 
confidence along 59 of the 102 sight lines surveyed.  For the highest 
quality sub-sample of the dataset, the high velocity detection frequency 
increases to 22 of 26 sight lines.  Forty of the 59 
sight lines have high velocity O\,VI $\lambda1031.926$ absorption
with  $W_\lambda > 100$ m\AA, and 27 have $W_\lambda > 150$~m\AA.  
Converting these O\,VI equivalent width detection frequencies
 into estimates of $N$(H$^+$) in the 
hot gas indicates
that $\sim60$\% of the sky (and perhaps as much as $\sim85$\%) 
is covered by hot ionized hydrogen at a level of 
$N({\rm H}^+)  \stackrel{_>}{_\sim} 8\times10^{17}$ cm$^{-2}$ if the high velocity gas has 
a metallicity similar to that of the Magellanic Stream ($Z\sim0.2-0.3$).
This detection frequency of 
hot H$^+$ associated with the  high velocity O\,VI appears to be
 larger than the value of $\sim37$\% found for high velocity warm gas with 
$N({\rm H\,I}) \sim 10^{18}$ cm$^{-2}$ traced through 21\,cm emission
(Lockman et al. 2002).

Some of the high velocity O\,VI is associated with well-known H\,I 
high velocity clouds (HVCs; see Wakker \& van~Woerden 
1997 for a review).  These include the Magellanic Stream, possibly
Complex~A, Complex~C, the Outer Arm, and several smaller H\,I clouds.
Some of the high velocity O\,VI features have no counterpart 
in H\,I 21\,cm emission.  These include discrete high velocity
features as well as broad positive-velocity
O\,VI absorption wings that blend with lower velocity O\,VI absorption
in the Galactic thick disk/halo. The 
discrete features may typify clouds located in the Local Group. 
The broad, high velocity O\,VI absorption wings are concentrated
mainly in the northern Galactic hemisphere  
and may trace either tidal debris or thick disk/halo gas 
that has been accelerated to high velocities by star-formation activity 
in the Galactic disk.

\subsection{Interpretation}

High velocity O\,VI is both widespread and common 
along complete paths through the Galactic halo.  The high velocity O\,VI 
traces numerous phenomena, including tidal interactions with the 
Magellanic Clouds (via the Magellanic Stream), accretion of low-metallicity
gas (e.g., Complex~C), highly ionized clouds (e.g., the Mrk~509 HVCs),
and the outflow of hot gas from the Galactic disk (e.g., the broad positive
velocity absorption features).  Distinguishing between all of the
possible phenomena occurring at large distances from the Galactic plane is 
absolutely essential for understanding the role of HVCs in Galactic 
evolution.

One possible explanation for some of the high velocity O\,VI is that
transition temperature gas arises at the boundaries between cool/warm 
clouds of gas and a very hot ($T > 10^6$\,K) Galactic corona or Local Group
medium.   Sources of the high velocity material might include infalling or 
tidally disturbed galaxies.  
Since viscous processes affect the gas but not the stars in interacting
systems, tracing the original source of the high velocity gas may prove 
difficult.  A hot, highly extended ($R > 70$ kpc)
corona or Local Group medium might be left over from the formation of the 
Milky Way or Local Group, or may be the result of continuous accretion of 
smaller galaxies over time.  
N-body simulations
of the tidal evolution and structure 
of the Magellanic Stream favor a low-density medium ($n < 10^{-4}$ cm$^{-2}$)
for imparting
weak drag forces to deflect some of the Stream gas and providing a possible
explanation for the absence of stars in the Stream (Gardiner 1999). 
Moore \& Davis (1994) also postulated a hot,
low-density corona to provide ram pressure stripping of some of the Magellanic
Cloud gas.
Hydrodynamical simulations of clouds moving through a hot, low-density 
medium show that weak bow shocks develop on the leading edges of the 
clouds as the gas is compressed and heated (Quilis \& Moore 2001).  
Even if the clouds are 
not moving at 
supersonic speeds relative to the ambient medium, some viscous or turbulent
stripping of the cooler gas likely occurs.

An alternative explanation for the O\,VI observed at high velocities 
may be that the clouds and any associated H\,I fragments are simply 
condensations within
large gas structures falling onto the Galaxy.
Cosmological structure formation models predict large numbers of cooling 
fragments embedded in dark matter, and some of these structures should be 
observable in O\,VI absorption as the gas pass through the 
$T=10^5-10^6$\,K
temperature regime.  Estimates of the number density of these structures 
are consistent with the observed IGM O\,VI detection rate (Tripp, Savage 
\& Jenkins 2000; Fang \& Bryan 2001; Savage et al. 2002a).  
This situation is in many ways analogous to the coronal
model described above because only about 30\% of the hot gas is detectable 
in O\,VI absorption, while the remaining $\sim70\%$ is too hot to 
observe  (Dav\'e et al. 2001).

The  tenuous hot Galactic corona or 
Local Group gas may manifest itself through X-ray absorption-line 
observations
of higher ionization species than O\,VI. For example, the 
amount of O\,VII in the hot gas
is given by $N$(O\,VII) = (O/H)$_\odot$ $Z$ f$_{O\,VII}$ $nL$,
where $Z$ is the 
metallicity of the gas, f is the ionization fraction, and $L$ is the 
path length.  At $T \sim 10^6$\,K, $f_{O\,VII} \approx 1$ 
(Sutherland \& Dopita 1993). For $n = 10^{-4}$ cm$^{-3}$,  
$N$(O\,VII) $\sim 2\times10^{16}~Z$~($L / 100$ kpc) (cm$^{-2}$).
Preliminary results 
(Fang et al. 2002; Nicastro et al. 2002) demonstrate
that O\,VII absorption is detectable near zero velocity at a level consistent 
with the presence of a large, nearby reservoir of hot gas.  A firm 
association of the
X-ray absorption with a hot Galactic corona or Local Group medium will
require additional observations since the ionization
mechanism and location of the higher ionization gas traced by O\,VII are
still uncertain (see Heckman et al. 2002).

\section{Looking Forward: The Need for a Large Space Telescope}

Studying the basic physical properties of gas in the vicinity of
galaxies is essential for understanding how galaxies evolve over time and 
quantifying the relationship between galaxies, the IGM,
and large-scale cosmic 
structures.  Key quantities to be determined include
 the elemental abundances, ionization state, and physical
properties ($n$, $T$, size) of the gas.  We have made a start at examining the 
properties of gas in the Milky Way and Local Group, but extending these 
studies to other groups of galaxies is required if we are to incorporate 
the local results into broader descriptions of galaxy
evolution and the formation of the  "cosmic web" of hot gas expected
in the local universe.  To make accurate assessments of the kinematics
and column densities of a wide range of ionization stages requires
high spectral resolution (R~$\sim20,000-50,000$), broad UV wavelength
coverage (preferably 900--3200\,\AA\ so that both zero and low-redshift 
systems could be observed in C\,III and O\,VI), and excellent 
sensitivity.

Current ultraviolet spectrographs (e.g., FUSE and STIS) are limited to 
observing QSOs with $m_B < 16$.  At this magnitude, the number 
of QSOs per square degree is $\ll 1$, and it is difficult to observe 
individual stars in galaxies other than the Milky Way and Magellanic 
Clouds.  The Cosmic Origins Spectrograph (COS), which is scheduled for 
installation in HST in 2004, will have about an order of magnitude greater 
sensitivity than STIS at far-UV wavelengths.  QSOs with $m_B \approx 17-18$
should be observable in reasonable integration times.  However, even at this 
greater sensitivity, the number of QSOs per square degree is still limited.
A large space telescope with a diameter of 4 meters (or better still, 8 meters)
equipped with a high efficiency spectrograph and detector could easily improve upon the COS sensitivity by a factor of 10 or more and dramatically 
increase the number of background sources available for spectroscopic\
studies.  At $m_B \approx 20$, the average separation between QSOs is 
only $\sim 7$\arcmin\ (see Shull et al. 1999), and there are numerous
possibilities for studying multiple sight lines through low-redshift galactic 
halos, groups, and clusters of galaxies.  Furthermore, at these 
sensitivities lightly reddened B supergiants 
could be observed out to distances of $\sim2$ Mpc, making it possible to 
study the gaseous content of Local Group galaxies such as M\,31 or M\,33 in 
unprecedented detail.

A 4\,m to 8\,m space telescope optimized for point source spectroscopy 
would provide an opportunity for an entirely new 
approach to studying the gaseous halos of galaxies and intragroup gas.  
Instead of our current methodology of blindly finding low-redshift 
absorption-line systems spectroscopically and 
then trying to identify (image) the galaxies responsible for the absorption,
one could target specific, well-studied galactic systems for spectroscopic
investigations of their gaseous content.  A similar pro-active approach may 
prove fruitful for studying the filamentary structures of the ionized IGM once 
they are identified through their O\,VI or X-ray emissions by
future observatories such as SPIDR or Constellation-X.
\smallskip
\\
\noindent
{\bf Acknowledgments:}
We thank Mike Shull and Ed Jenkins for stimulating 
discussions and comments on the interpretation of the high velocity O\,VI
properties.  

\references
\reference{}Anders, E., \& Grevesse, N.  1989, Geochim. Cosmochim. Acta., 
        53, 197
\reference{}Dav\'e, R., Cen, R., Ostriker, J.P., et al. 2001, ApJ, 552, 473
\reference{}Fang, T.T., \& Bryan, G.L.  2001, ApJ, 561, L31
\reference{}Fang, T., Sembach, K.R., \& Canizares, C.R. 2002, ApJ,
	in preparation
\reference{}Gardiner, L.T., 1999, in ``The Stromlo Workshop on High 
Velocity Clouds'', ASP Conf. 166, eds. B.K. Gibson \& M.E. Putman,
        (San~Francisco: ASP), 292
\reference{}Grevesse, N., \& Noels, A. 1993, in Origin of the Elements, ed. 
	N. Prantzos, E. Vangioni-Flam, \& M. Cass\'e, (Cambridge: Univ. Press), 15
\reference{}Heckman, T.M., Norman, C.A., Strickland, D.K., \& Sembach, K.R.
	2002, ApJ, in press [astro-ph/0205556]
\reference{}Lockman, F.J., Murphy, E.M., Petty-Powell, S., \& Urick, V. 2002,
	ApJS, in press [astro-ph/0201039]
\reference{}Moore, B., \& Davis, M.  1994, MNRAS, 270, 209
\reference{}Morton, D.C.  1991, ApJS, 77, 119
\reference{}Nicastro, F., Zezas, A., Drake, J., et al.  2002, ApJ, 573, 157
\reference{}Quilis, V., \& Moore, B.  2001, ApJ, 555, L95
\reference{}Savage, B.D., Sembach, K.R., Tripp, T.M., \& Richter, P.  2002a, 
	ApJ 564,  631
\reference{}Savage, B.D., Sembach, K.R., Wakker, B.P., et al. 2002b, 
	ApJS, submitted [astro-ph/0208140]
\reference{}Sembach, K.R., Wakker, B.P., Savage, B.D., et al. 2002, 
	ApJS, submitted [astro-ph/0207562]
\reference{}Shull, J.M., Savage, B.D., Morse, J.A., et al.  1999, 
	``The Emergence of the Modern Universe'', a white paper of the 
	UV-Optical Working Group
\reference{}Sutherland, R.S., \& Dopita, M.A.  1993, ApJS, 88, 253
\reference{}Tripp, T.M., Savage, B.D., \& Jenkins, E.B.  2000, ApJ 534, L1
\reference{}Verner, D.A., Barthel, P.D., \& Tytler, D.  1994, A\&AS, 108, 287
\reference{}Wakker, B.P., Savage, B.D., Sembach, K.R., et al. 2002, 
	ApJS, submitted [astro-ph/0208009]
\reference{}Wakker, B.P., \& van~Woerden, H.  1997, ARA\&A, 35, 217

\end{document}